\documentclass[aps,prl,reprint,superscriptaddress,twocolumn]{revtex4}
\usepackage[dvips]{graphicx}
\usepackage{subfigure}
\usepackage{bm}

\begin{document}

\title{ Spinmotive Force due to Intrinsic Energy of Ferromagnetic Nanowires }

\author{Y. Yamane}
\affiliation{Advanced Science Research Center, Japan Atomic Energy Agency, Tokai 319-1195, Japan}
\affiliation{Institute for Materials Research, Tohoku University, Sendai 980-8577, Japan}
\author{J. Ieda}
\affiliation{Advanced Science Research Center, Japan Atomic Energy Agency, Tokai 319-1195, Japan}
\affiliation{CREST, Japan Science and Technology Agency, Tokyo 102-0075, Japan}
\author{J. Ohe}
\affiliation{CREST, Japan Science and Technology Agency, Tokyo 102-0075, Japan}
\affiliation{Department of Physics, Toho University, Funabashi, 274-8510 Japan}
\author{S. E. Barnes}
\affiliation{Physics Department, University of Miami, Coral Gables, Florida 33124, USA}
\author{S. Maekawa}
\affiliation{Advanced Science Research Center, Japan Atomic Energy Agency, Tokai 319-1195, Japan}
\affiliation{CREST, Japan Science and Technology Agency, Tokyo 102-0075, Japan}

\begin{abstract}

We study, both analytically and numerically, a spinmotive force arising from inherent magnetic energy of a domain wall in a wedged ferromagnetic nanowire.
In a spatially-nonuniform nanowire, domain walls are subjected to an effective magnetic field, resulting in spontaneous motion of the walls.
The spinmotive force mechanism converts the ferromagnetic exchange and demagnetizing energy of the nanowire into the electrical energy of the conduction electrons through the domain wall motion.
The calculations show that this spinmotive force can be several microvolts, which is easily detectable by experiments.

\end{abstract}

\maketitle

%
%

Recently, an electromotive force of spin origin has been theoretically predicted\cite{barnes1} and experimentally observed\cite{yang,tanaka}.
This ``spinmotive force'' reflects the energy conversion of magnetic energy of a ferromagnet into electric energy of conduction electrons via the spin exchange interaction.
This offers a new functionality for future spintronic devices.
Theoretically, it has been pointed out that the spinmotive force is produced by the spin electric field acting on conduction electrons\cite{volovik,sanvito,duine,tser,yuta}:
\begin{equation}
\bm{E}_\mathrm{s}=-\frac{P\hbar}{2e}\bm{m}\cdot\left(\partial_t\bm{m}\times\nabla\bm{m}\right)
\label{field}\end{equation}
where $\bm{m}$ is the unit vector paralell to the direction of the local magnetization in a ferromagnet, $P$ the spin polarization of the conduction electrons, and $e$ the elementaly charge.
It is requied that the magnetization depends both on time and space.
To demonstrate the spinmotive force, Yang \textit{et al.} studied a field-induced domain wall (DW) motion in a uniform nanowire\cite{yang}, in which the moving DW releases its Zeeman energy into conduction electrons through the spinmotive force mechanism.
On the other hand, it has been suggested that a DW motion can be induced solely by ``shape effect''\cite{barnes2,ieda}.
In addition to the Zeeman energy, a DW has a surface-tension-like intrinsic energy due to the ferromagnetic exchange and magnetic anisotropic energy of the magnetization.
Thus in a spatially-nonuniform nanowire a DW spontaneously tends to move to narrower regions in order to reduce its intrinsic energy, which is proportinal to the cross-sectional area of the wire\cite{ieda,barnes2}.
The shape-effect-induced DW motion thereby gives rise to a spinmotive force without any external field, using the inherent energy of the magnetic texture, i.e., the ferromagnetic exchange and anisotropic energy\cite{barnes2}.
In reality, the DW motion in a spatially-nonuniform nanowire shows highly nonlinear behaviors\cite{ieda,kyoto1,kyoto2,parkin,barnes3}, for a quantitative understanding of which numerical methods are of particular importance.

In this paper, we study, both analytically and numerically, the spinmotive force in a wedged magnetic nanowire.
The results of the numerical simulations strongly support this concept, and show that this spinmotive force can be large and lasts long enough to be detected in experiments.

\begin{figure}[b]
\begin{center}
\includegraphics[width=85mm]{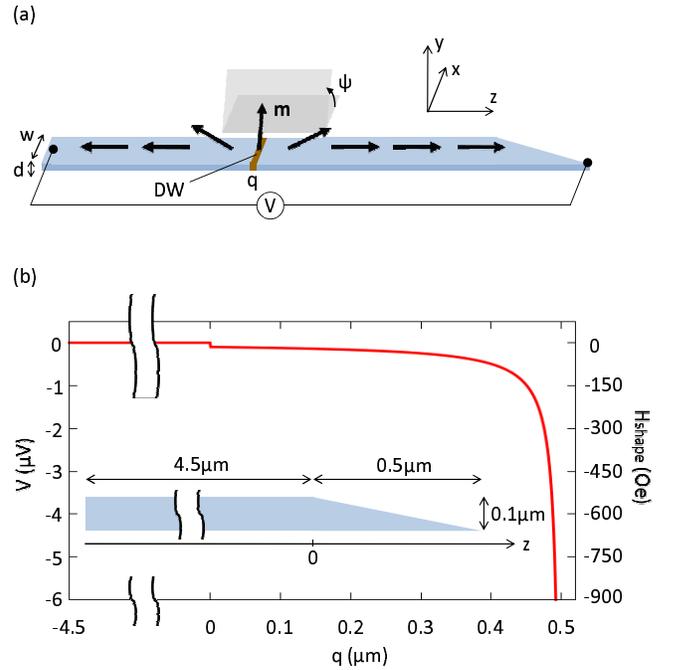}
\end{center}
        \caption{ (a) The schematic illustration of a Ni$_{81}$Fe$_{19}$ wedged nanowire containng a single DW.
                      The arrows indicate the magnetizaiton direction $\bm{m}$.
                      The DW is characterized by two collective coordinates, i.e., $q$, the center position, and $\psi$, the tilt angle from the easy plane. 
                  (b) The spinmotive force and the shape-effect magnetic field, calculated by Eqs.~(\ref{shape}), (\ref{psi}), (\ref{h}), and (\ref{v}), with $H=0$ and the tail-to-tail configuration, as a function of the DW position $q$.
                      The inset describes the dimensions of the sample.
                }
\label{f1}
\end{figure}

%
%

We consider a Permalloy thin wire in which one of its ends is wedged as illustrated in Fig.~\ref{f1}.
The length, width, and thickness of the wire are $5$ $\mu$m, $w=100$ nm, and $d=4$ nm, respectively.
The cross-sectional area of the wire $A$ tapers toward the end as 
\begin{equation}
A(z)  =  \left( - \frac{1}{5}z + w \right) d.
\label{shape}\end{equation}

First, we analyticaly investigate influence of the wedge on the spinmotive force based on a collective-coordinate model.
Here we employ the one-dimensional magnetization dynamics along the $z$-axis, assuming uniformity in the lateral direction (the $x$-$y$ plane), in which the dynamics of a DW is characterized by two collective coordinates, i.e., the center position $q$ and the tilt angle $\psi$ of the DW plane relative to the $x$-$z$ plane as indicated in Fig.~\ref{f1} (a).
The time evolution of ($q,\psi$) is described by the reduced Landau-Lifshiz-Gilbert equations\cite{ieda}
\begin{eqnarray}
\dot{q}     &=&  \mp \frac{ \Delta \gamma } { 1 + \alpha^2 }  \left[  \alpha \left( H + H_\mathrm{shape} \right)  +         \frac {H_K} {2} \sin 2 \psi   \right],    \label{q}\\
\dot{\psi}  &=&      \frac{ \gamma }        { 1 + \alpha^2 }  \left[                H + H_\mathrm{shape}          -  \alpha \frac {H_K} {2} \sin 2 \psi   \right],    \label{psi}
\end{eqnarray}
where $\gamma$ is the gyromagnetic ratio, $\alpha$ is the Gilbert damping constant, $\Delta$ is the wall width parameter, $H$ is an external magnetic field, and $H_K$ is a perpendicular anisotropy field due to the demagnetizing interaction.
Here and hereafter, upper (lower) sign corresponds to the tail-to-tail (head-to-head) configuration of the DW.
In a spatially-nonuniform nanowire, in addition to $H$, DWs are subjected to the ``shape-effect'' field\cite{barnes2,ieda}:
\begin{equation}
H_\mathrm{shape} = \pm \frac{\sigma}{2M_\mathrm{s}}\frac{\partial}{\partial q}\ln A(q),
\label{h}\end{equation}
where $\sigma$ is surface tension energy of a DW and $M_\mathrm{s}$ is the saturation magnetization.
As in the text book\cite{text}, energy minimization gives $\sigma=\sigma_0\left(1+Q^{-1}\sin^2\psi \right)^{1/2}$.
Here $\sigma_0=\sqrt{A_\mathrm{s}K_\mathrm{u}}$ with $A_\mathrm{s}$ the exchange stiffness constant, $K_\mathrm{u}$ the uniaxial anisotropy constant, and $Q$ is the ratio of $K_\mathrm{u}$ to the perpendicular anisotropy constant.
This illustrates that the gradient of the intrinsic DW energy $\sigma A(q)$ exerts a force acting on the DW.
In the present system, the magnetic anisotropy comes from the demagnetizing field while the crystalline magnetic anisotropy is negligible.
It is the exchange and the demagnetizing interaction between the magnetization that is responsible for $H_\mathrm{shape}$.

From Eqs.~(\ref{q}) and (\ref{psi}), in a spatially-nonuniform magnetic nanowire, $H_\mathrm{shape}$ can drive a DW and make precess the DW plane even without the external field $H$.
In the light of this simplification, with some sign convention the spinmotive force across the DW is obtained as
\begin{equation}
V  =  \int dzE_\mathrm{s}  =  \pm \frac{P\hbar}{e} \dot{\psi}.
\label{v}\end{equation}
Equation~(\ref{v}) indicates that the sign of the spinmotive force reflects that of the precession rate of the DW plane $\dot{\psi}$, and is independent of the translational motion of the DW\cite{barnes1,yang}.
When $ \left| H + H_\mathrm{shape} \right| $ is smaller than the Walker breakdown field $H_\mathrm{W}$, the Larmor torque due to $ H + H_\mathrm{shape} $ cannot overcome the countertorque due to the perpendicular anisotropy; 
$ H + H_\mathrm{shape} $ in Eq. (\ref{psi}) are canceled out by the third term with a certain $\psi$, resulting in $\dot{\psi}=0$.
In contrast, if $ \left| H + H_\mathrm{shape} \right| $ is larger than $H_\mathrm{W} = \alpha H_K/2$, the DW plane precesses with finite $\dot{\psi}$.
Thus, the spinmotive force appears only when $ \left| H + H_\mathrm{shape} \right| $ is above the Walker breakdown field.
Once the DW configuration is fixed, the sign of the spinmotive force is determined by the direction of $ H + H_\mathrm{shape} $.

The shape-effect field $H_\mathrm{shape}$ and the spinmotive force $V$, calculated from Eqs.~(\ref{shape}), (\ref{psi}), (\ref{h}) and (\ref{v}), with $H=0$ and the tail-to-tail configuration, as a function of the DW position $q$ are shown in Fig.~\ref{f1} (b).
Typical values for a Permalloy nanowire are employed; $\gamma=1.76\times 10^{11}$ Hz/T, $\alpha=0.01$, $M_\mathrm{s}=1$ T, $A_\mathrm{s}=1.3\times 10^{-11}$ J/m, $K_\mathrm{u}=10^5$ J/m$^3$, and $P=0.6$.
We neglect the third term in Eq.~(\ref{psi}), in which $\alpha H_K$ is tens of Oersted for the Permalloy nanowire, and $\sigma\simeq\sigma_0$ is assumed.
In this scheme $V$ is proportional to $H_\mathrm{shape}$, and in the wedged part ($z>0$) the finite spinmotive force appears.
Figure~\ref{f1} (b) indicates that the spinmotive force of several microvolts can be obtained in the present system without any applied field but using the intrinsic magnetic energy of the nanowire.

%
%

\begin{figure}[t]
\begin{center}
\begin{tabular}{cc}
\includegraphics[width=85mm]{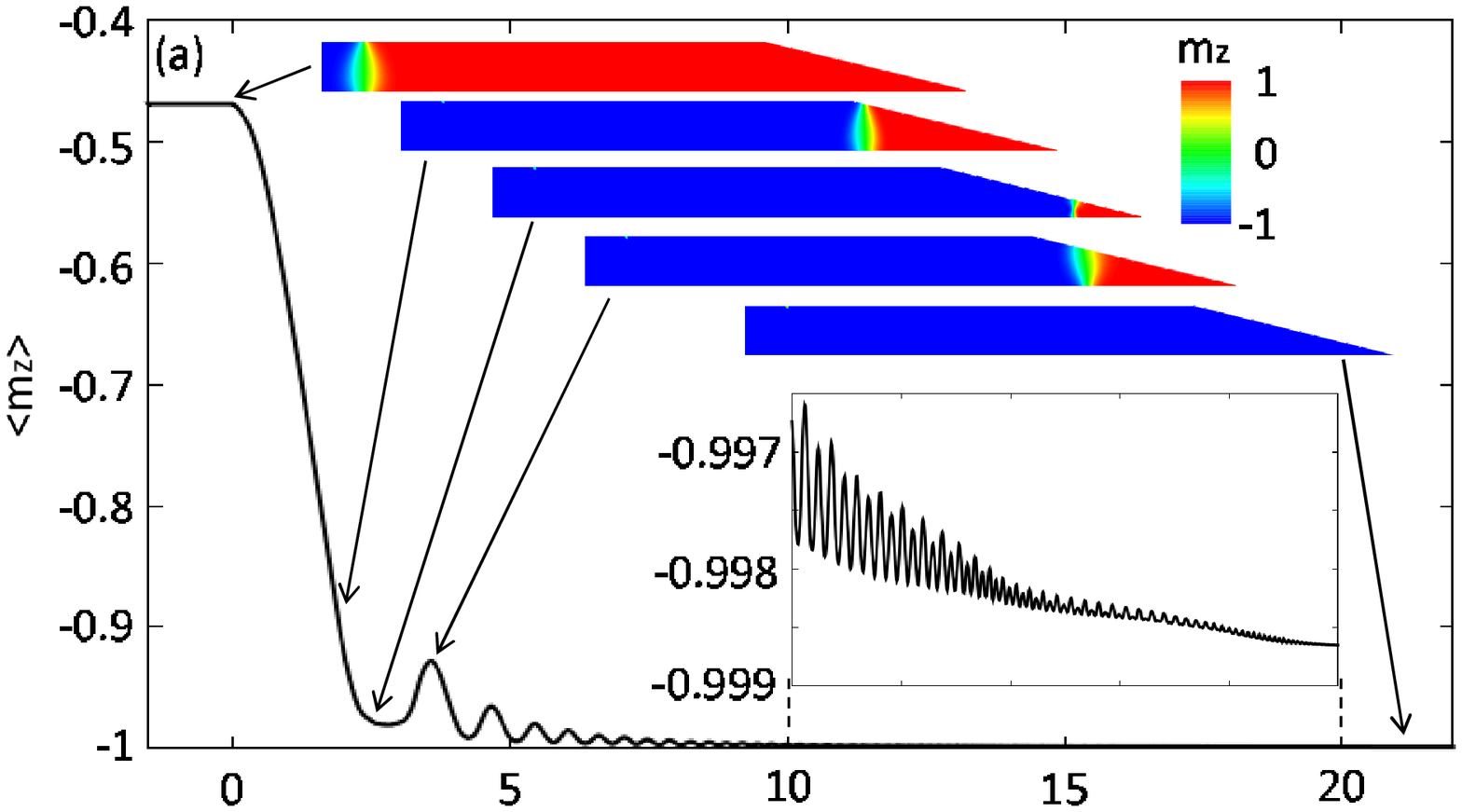}\\
\includegraphics[width=85mm]{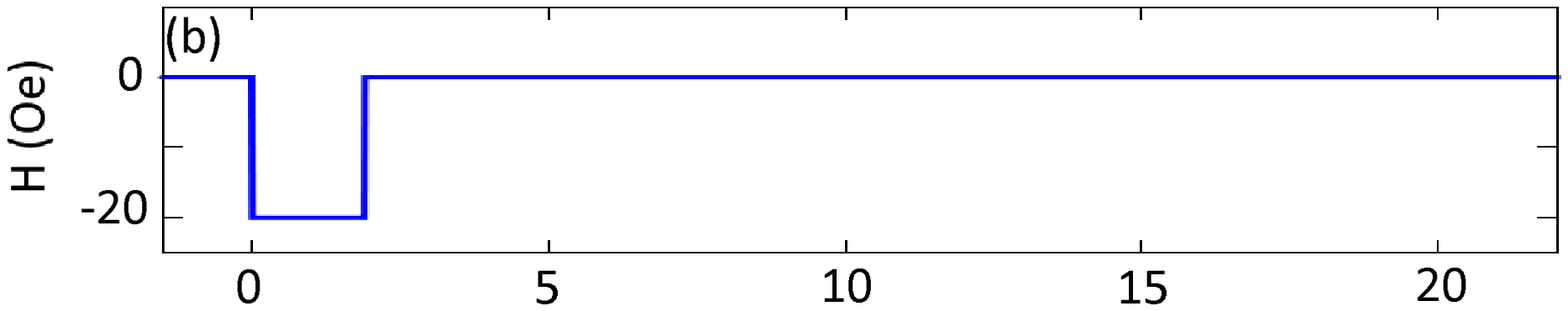}\\
\includegraphics[width=85mm]{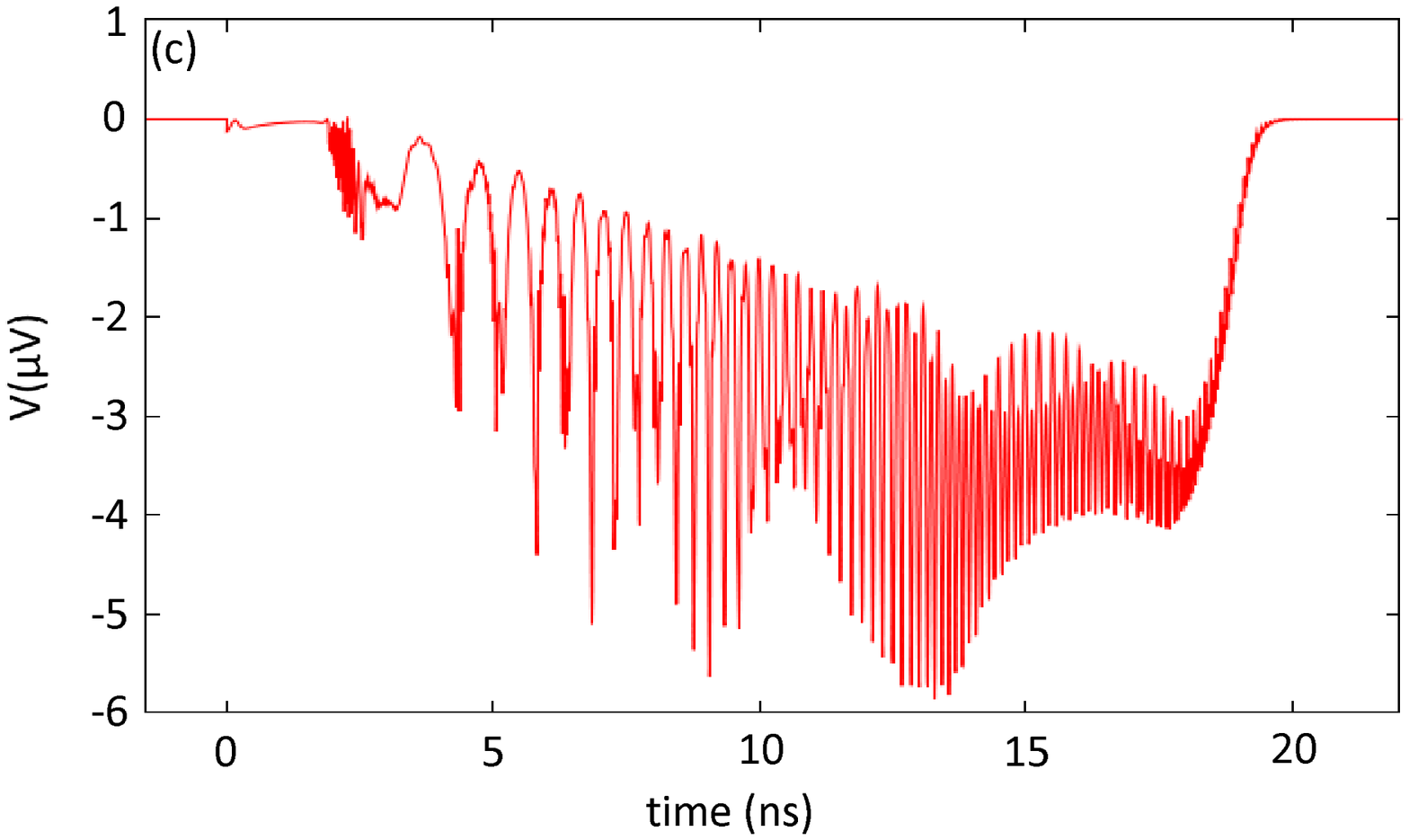}
\end{tabular}
\end{center}
        \caption{ The time evolution of (a) the spatially averaged $z$-component of the magnetization $\left\langle m_z\right\rangle$ (the inset showing a magnified image) with the five snapshots of the distribution of $m_z$ corresponding to time indicated by the arrows, (b) the applied magnetic field, and (c) the spinmotive force between the ends of the nanowire.
                }
\label{f2}
\end{figure}

The collective-coordinate model imposes a number of assumptions, i.e., the uniformity in the $x$-$y$ plane , a rigid wall structure, and constant magnetic shape anisotropies.
In fact, these assumptions are no longer guaranteed without check for nanowires with the shape modulation and in particular for relatively higher driving field region ($\left| H+H_\mathrm{shape} \right| \gg H_\mathrm{W}$).
To confirm the above prediction and to obtain the detailed real-time spectrum of the induced spinmotive force with high accuracy, here we perform numerical calculations free from the constraint.
We solve the original Landau-Lifshiz-Gilbert equation
\begin{equation}
\dot{\bm{m}} \left( \bm{r}, t \right)  =   -  \gamma \bm{m} \left( \bm{r}, t \right) \times \bm{H}_\mathrm{eff} + \alpha \bm{m} \left( \bm{r}, t \right) \times \dot{\bm{m}}\left(\bm{r},t\right),
\label{llg}\end{equation}
using the OOMMF code\cite{oommf} with dividing the sample into $4\times 4\times 4$ nm$^3$ cells.
Here $\textrm{\boldmath $H$}_\mathrm{eff}$ is the effective magnetic field including the external, exchange and demagnetizing fields.
The parameters are the same as used in the one-dimensional calculation.
Here the demagnetizing field is taken into account by the dipole interaction, instead of $K_\mathrm{u}$.
To track the DW position we calculate the time evolution of spatially-averaged value of the $z$-component of the magnetization, $\left\langle m_z \right\rangle = ( 1 / V ) \int m_z dV$, as shown in Fig.~\ref{f2} (a).
Initially ($t\leq 0$), a transverse tail-to-tail DW is trapped at $z=-1$ $\mu$m, where the $4\times 4\times 4$ nm$^3$ defect is introduced.
From $t=0$ to $1.9$ ns a constant magnetic field pulse $\bm{H}=(0,0,H)$, where $H=-20$ Oe, is applied, see Fig.~\ref{f2} (b); 
during this period, the DW has been dislodged from the defect, traveled about 1 $\mu$m and entered the wedged part.
The DW maintains its transverse structure through the field-induced motion.
Even after the applied field is turned off at $t=1.9$ ns, the DW keeps moving forward and slides down the slope with the help of the shape-effect.
In this stage there arises an oscillation of the DW position, i.e. the back and forth motion of the DW, with the periodic nucleation and annihilation of antivortices.
This is because of the Walker breakdown induced by the shape-effect.
In the one-dimensional model, as the tilt angle $\psi$ develops through the Walker breakdown motion, $(\Delta\gamma H_K/2)\sin 2 \psi$ in Eq.~(\ref{q}) oscillates and thus $\dot{q}$ can be negative.
This oscillatory motion continues until the DW reaches the end of the nanowire and disappears at $t\simeq 20$ ns.
We remark here that after the whole event, the exchange and the demagnetizing energy of the nanowire has obviously decreased.
The spinmotive force mechanism, which guarantees the total energy conservation, dictates that the decreased magnetic energy is received by the conduction electrons, resulting in an electromotive force.

A measurable voltage between given two points can be calculated as the spatial difference of the spin-independent electric potential $\phi(\bm{r},t)$ between the points, which is obtained by solving the Poisson equation $\nabla^2\phi(\bm{r},t)=-\nabla\cdot\bm{E}_\mathrm{s}(\bm{r},t)$\cite{yang,ohe}.
Figure~\ref{f2} (c) shows the spinmotive force between the ends of the nanowire [see Fig.~\ref{f1} (a)] versus time.
In spite of the field-induced DW motion, from $t=0$ to $1.9$ ns, a very small spinmotive force is observed since $\left|H\right|=20$ Oe is below the Walker breakdown field for the present nanowire.
After the system goes into the Walker breakdown, with no applied magnetic field, the finite spinmotive force is induced until the DW vanishes at $t \simeq 20$ ns.
The observed spinmotive force is of the order of microvolts, oscillating but unipolar.
These facts agree with the analytical prediction from Eq.~(\ref{v}).
The oscillation of the spinmotive force reflects the DW structure transformation.
Apparently $\dot{\psi}$ becomes larger in an anti-vortex motion than in that of a transeverse wall and so does $V$.
The peaks of the spinmotive force signal in Fig.~\ref{f2} (c) are associated with the traverse motion of an antivortex, i.e., the plateau in the DW position curve in Fig.~\ref{f2} (a).
The DW moves back and forth with changing its structure, whereas, the sign of the spinmotive force is unchanged and determined by the direction of the shape-effect magnetic field as estimated in Fig.~\ref{f1} (b) [see also the discussion below Eq.~(\ref{v})].

%
%

In conclusion, we have investigated the spinmotive force in a wedged magnetic nanowire, and numerically confirmed the spinmotive force of several microvolts for tens of nano seconds, using only the inherent magnetic energy of the nanowire.
The behavior of the spinmotive force depends on the shape and the material nature of the sample, which implies the posibility of various applications.

%
%

\acknowledgments
We are grateful to M. Hayashi and S. Mitani in National Institute for Material Science for helpful comments on this work.
This research was suported by Giant-in-Aid for Science Research from MEXT, Japan, and the Next Generation Supercomputer Project, Nanoscience Program from MEXT, Japan.

%
%


\begin{thebibliography}{99}


\bibitem{barnes1}  S. E. Barnes and S. Maekawa: Phys. Rev. Lett. {\bf{98}} (2007) 246601.
\bibitem{yang}     S. A. Yang, G. S. D. Beach, C. Knutson, D. Xiao, Z. Zhang, M. Tsoi, Q. Niu, A. H. MacDonald, and J. L. Erskine: Phys. Rev. B {\bf{82}} (2010) 054410.
\bibitem{tanaka}   P. N. Hai, S. Ohya, M. Tanaka, S. E. Barnes, and S. Maekawa: Nature {\bf{458}} (2009) 489.
\bibitem{volovik}  G. E. Volovik: J. Phys. C {\bf{20}} (1987) L83.
\bibitem{sanvito}  M. Stamenova, T. Todorov, and S. Sanvito: Phys. Rev. B {\bf{77}} (2008) 054439.
\bibitem{duine}    R. A. Duine: Phys. Rev. B {\bf{77}} (2008) 014409.
\bibitem{tser}     Y. Tserkovnyak and M. Mecklenburg: Phys. Rev. B {\bf{77}} (2008) 134407.
\bibitem{yuta}     Y. Yamane, J. Ieda, J. Ohe, S. E. Barnes, S. Maekawa: J. Appl. Phys. $\bm{109}$ (2011) 07C735.
\bibitem{barnes2}  S. E. Barnes, J. Ieda, and S. Maekawa: Appl. Phys. Lett. {\bf{89}} (2006) 122507.
\bibitem{ieda}     J. Ieda, H. Sugishita, and S. Maekawa: J. Magn. Magn. Mat. {\bf{322}} (2010) 1363.
\bibitem{kyoto1}   A. Yamaguchi, T. Ono, S. Nasu, K. Miyake, K. Mibu, and T. Shinjo: Phys. Rev. Lett. {\bf{92}} (2004) 77205.
\bibitem{kyoto2}   A. Himeno, K. Kondo, H. Tanigawa, S. Kasai, and T. Ono: J. Appl. Phys. {\bf{103}} (2008) 07E703.
\bibitem{parkin}   S. S. P. Parkin, M. Hayashi, and L. Thomas: Science {\bf{320}} (2008) 190.
\bibitem{barnes3}  V. Lecomte, S. E. Barnes, J. P. Eckmann, and T. Giamarchi: Phys. Rev. B {\bf{80}} (2009) 054413.
\bibitem{text}     A. P. Malozemoff and J. C. Slonczewski: \textit{Magnetic Domain Walls in Bubble Materials} (Academic Press, New York, 1979) Chap. 5, 7, and 10.
\bibitem{oommf}    http://math.nist.gov/oommf/
\bibitem{ohe}      J. Ohe and S. Maekawa: J. Appl. Phys. {\bf{105}} (2009) 07C706.


\end{thebibliography}
\end{document}